 \documentstyle[prl,aps]{revtex}
\input epsf.tex

\begin{document}

\preprint{8-May-96 RL DRAFT-1}

\draft

\twocolumn[

\title{Quantum Logical Operations on Encoded Qubits}

\author{Wojciech Hubert Zurek
and Raymond Laflamme }
 
\address{\vspace*{1.2ex}
 	\hspace*{0.5ex}{Theoretical Astrophysics, T-6, MS B288}\\
 	 Los Alamos National Laboratory, Los Alamos, NM 87545, USA}

\date{\today}
\maketitle

]


\begin{abstract}
 We show how to carry out quantum logical operations 
(controlled-not and Toffoli gates) on encoded qubits for several 
encodings which protect against various 1-bit errors.
This improves the reliability of these operations by allowing one to 
correct for one bit errors which either preexisted  or occurred in 
course of operation.  The logical operations we consider allow one 
to cary out the vast majority of the steps in the quantum factoring 
algorithm.  Thus, our results help bring
quantum factoring and other quantum computations closer to reality
\end{abstract}





Schemes for encoding individual quantum bits into ``qubytes'' consisting 
of several qubits were recently proposed \cite{shor95,steane,calde,LMPZ,bennett}
and shown to offer a significant measure of protection against the 
environment-induced decoherence\cite{Zurek91,Landauer95,Unruh94,CLSZ95} 
and other possible sources of errors. However, as pointed by Shor\cite{shor95}, 
who devised first such encoding, 
the usefulness of this strategy in the context of quantum computation is limited
as long as -- for the purpose of carrying out logical operations -- one would 
need to ``decode''
the qubit and use it in its ``bare'' form to compute.  Here we present the 
first implementation of logical gates on encoded qubits and evaluate their 
efficiency.

We limit our presentation, for simplicity, to the linear codes proposed by Steane\cite{steane}.  He has devised two such encodings,
the first of which protects only against decoherence
\begin{eqnarray}
|0_L\rangle = |000\rangle + |011\rangle + |101\rangle + |110\rangle
\nonumber \\
|1_L\rangle = |111\rangle + |100\rangle + |010\rangle + |001\rangle \ ,
\end{eqnarray}
while the second, 7-bit code is the shortest {\it linear} code which is capable of decoding with general 1-bit errors:
\begin{eqnarray}
|0_L &\rangle & = |0000000\rangle + |1010101\rangle + |0110011\rangle + |1100110\rangle  \nonumber\\
&+& |0001111\rangle + |1011010\rangle + |0111100\rangle + |1101001\rangle 
\nonumber \\
|1_L &\rangle & = |1111111\rangle + |0101010\rangle + |1001100\rangle + |0011001\rangle \nonumber\\
& +& |1110000\rangle + |0100101\rangle + |1000011\rangle + |00101101\rangle \ .
\nonumber \\
\end{eqnarray}
Any linear combinations of these logical states is also part of the code\cite{steane}.
We will, later on, utilize other natural logical states such as; 
\begin{equation}
|\pm_L\rangle \  = \  |0_L\rangle \pm |1_L\rangle \  = \ |\pm\pm\pm\rangle
\end{equation}
for the 3-bit code.

Three different implementations of the controlled not ({\tt CNOT}) for the 3-bit code
are shown in Figure 1 as examples of many more we have devised and will
discuss elesewhere\cite{zulaf}.  It is easiest to start  the discussion with 
the encoding of Fig.1a.  The reason it works is simple to understand from the structure of the logical $|0_L\rangle$ and $|1_L\rangle$ in Eq.(1):
the logical $|0_L\rangle$ has all the possible states with an even number of 1's while
$|1_L\rangle$ has all the states with an odd number of 1's. Therefore 
by flipping any bit we transform one of the logical states into its logical opposite.
The gate of Fig.1a will
simply flip the qubits in the target qubyte when the control qubit will
be in the state 1.  There will be an even number of such flips if the control
is in $|0_L\rangle$ but an odd number of flips for the control qubyte $|1_L\rangle$.  Consequently:
\begin{equation}
(\alpha|0_L^c\rangle+\beta|1_L^c\rangle|Q^t\rangle
\stackrel{{\tt CNOT}}{\Longrightarrow}
\alpha|0_L^c\rangle|Q^t\rangle + \beta|1_L^c \rangle|\neg Q^t\rangle \ .
\end{equation}
Above, superscipts ``{\it c}'' and ``{\it t}'' designate the ``control'' and ``target'' qubytes respectively, while;
\begin{eqnarray}
 |\neg Q^t\rangle = a|\neg 0_L^t\rangle + b|\neg 1_L^t\rangle
= a|1_L^t\rangle + b|0_L^t\rangle \ .
\end{eqnarray}

A similar explanation demonstrates the action of the {\tt CNOT} of Fig.1b: 
As the only thing that matters is whether there is an even or odd number of 1's in the 
target qubyte, one might as well operate on the top qubit only.  If the control bit has an odd number of 1's (because it encodes $|1_L^c\rangle$), the state of the top qubit will be flipped. Therefore, the logical state of the target qubyte as a whole will change.  By contrast $|0_L^c\rangle$ will result in an even number of flips of the top qubit, which means that its state remains unaffected.

It is important  to point out that in these two schemes (Fig.1a and Fig.1b) the target byte 
remains ``in the code'' inbetween the individual 1-bit {\tt cnot}'s.  Thus, one can also apply Steane's error correction scheme inbetween.

The last diagram in Fig.1c shows a still different implementation
of the {\tt CNOT}.  The gate used there affects a flip of the top bit in the target 
qubyte depending on the sum (modulo 2) of the state of the control qubyte.
In a sense, this last implementation  summs up the essence of our collective {\tt CNOT}: to carry it out one needs to implement a flip of any
single qubit in the target qubyte if the control qubyte has an odd number
of 1's.  

This last design for the collective {\tt CNOT} may also have the advantage
of a straightforward implementation in at least one of the proposed realizations
of a quantum computer -- the linear trap computer\cite{Cirac95}.
There one can imagine three copies of the memory coexisting in a single trap,
and sharing the single ``bus'' phonon.   That phonon can then be used as a
target bit of three  {\tt cnot} operations with the three relevant qubits (one from
the memory of each of the three parallel copies) acting as a control.  The state of the phonon will then store the information  about the parity of the control qubyte, 
and can be used as a control qubit to affect the state of one of the bits of 
the target qubyte, completing the design of Fig.1c.  The disadvantage of this implementation is that the phonon which acts as an ``ancilla'' qubit is not 
protected but the simplicity of the design
may prove to favor it anyway, especially since the phonon bus bit 
can be stabilized using the watchdog effect method\cite{Zurek84}. 

An essentially identical strategy works for the Steane 7-bit code, Eq.(2).
Again, the logical zero is even in the number of 1's and logical one is odd, so the {\tt CNOT} shown in Fig.2 (which is of course direct analog of Fig.1a) will do the job.
The strategy can also be applied to the 5 bit code 
we have previously proposed \cite{LMPZ}.  It suffices to know
that the first, second and fifth bit give the parity 
which distinguish the logical state
in this code.  This observation can be supplemented with the 
fact that in that 5 bit code
the $|1_L\rangle$ is obtained from the $|0_L\rangle$ by flipping the first bit
and changing the sign if it was a 1, to
create a {\tt CNOT} circuit without decoding the 5 bits.  The details
will be shown elsewhere\cite{zulaf,pacs96}.

{\tt CNOT} is a very useful logical operation, but it is not classically 
universal\cite{toffoli}.  That is, one cannot build a universal classical computer 
using only {\tt CNOT}'s.  More sophisticated logical gates are required for 
that 
purpose.  The Toffoli gate ({\tt T}-gate) is an example of a universal reversible 
logical gate: it can be used to implement a general purpose classical computer.
The {\tt T}-gate has two control bits both of which have to be ``1'' if the 
target qubit is to be flipped.
We show now how it can be implemented 3-bit qubytes of Eq.(1).

The {\tt T}-gate cannot be of course implemented using only {\tt CNOT}'s: If that was possible, 
{\tt CNOT} itself would be universal, which is not the case.  One additional 
operation which we shall require is a ``square root'' of controlled not (called $V$).
That is, if $U$ is the action of {\tt CNOT} on the target bit in the case when 
the control bit is unity;
\begin{equation}
V^2= U \ .
\end{equation}
This definition does not constrain $V$ uniquely but only up to a unitary rotation.  
A possible form of $V$ is 
\begin{equation}
V= \frac{1}{\sqrt{2}}\left (\begin{array}{cc} 1& i \\ i& 1 \end{array}\right) \ .
\end{equation}

With the help of $V$ and other logical operations we have already introduced Toffoli
 gate can be implemented through the design shown in Fig.3, which is 
related to the quantum {\tt T}-gate design of Barenco et al.\cite{barenco}.
The key to understanding this design is Fig.1b, which shows that in 
order to convert a logical state to its negation in Steane's code, Eq.(1), 
one can work with just one qubit.  In the implementation of Fig.3 we have elected to work with the ``top'' qubit of the target qubyte.  Now, when the control qubit CI is in a logical state 1 (0), the operation $V$ (or identity ${\cal I}$) will 
be carried out on the top qubit.  The three {\tt cnot}'s then compute the 
sum (mod 2) 
of CI and CII. If that sum is 0 (which it is whenever CI and CII are the same)
the operation ${\cal I}$ ($V$)
will be carried out.  The three subsequent  {\tt cnot}'s restore qubytes CII to its 
original logical state.  Thereafter, the operation $V$ (${\cal I}$) is carried out
depending on the state of CII.  The net effect for all the possibilities is:

\begin{center}
\hskip 1.7 truein\vbox{\tabskip=0pt \offinterlineskip
\def\tablerule{\noalign{\hrule}}
\halign to 125    pt
     {\strut#&\vrule#\tabskip=0pt& \vrule # &\vrule
      \hfil #  \vrule  \tabskip=0pt\cr\tablerule
& & \hfil    &  \cr
&\ \ CI &  \ \  CII  &\ \ Operation on\ \cr
& & \hfil $ $\hfil    & \hfil \ the target byte
      \cr\tablerule 
&\hfil 0\hfil& \hfil 0   \hfil &  ${\cal I}$ \ \ \ \ 
      \cr\tablerule 
&\hfil 1\hfil& \hfil 0   \hfil & $VV^\dagger={\cal I} $ \ \ 
      \cr\tablerule 
&\hfil 0\hfil& \hfil 1   \hfil & $V^\dagger V={\cal I}$  \ \ 
      \cr\tablerule 
&\hfil 1\hfil& \hfil 1   \hfil & $V  V=U$ \ \ 
      \cr\tablerule 
\cr}}
\end{center}
\begin{center}
\vskip -0.25truein
Table 1. Truth table for the quantum Toffoli gate.
\end{center}

Consequently, the above design accomplishes the action of a Toffoli gate.  
Moreover, the two control bytes are always in the code and can be 
intermittently corrected.  By contrast, the target qubyte is not in 
the code while the operations $V$ and $V^\dagger$ are taking place, 
but that might not be a major problem, as it can be corrected for 
both immediately before and after the gate.
Moreover, only its top qubit is used and that can happen no more than 
three times -- much less than the correctable qubytes CI and CII.

Although {\tt T}-gates are not universal for quantum computer (they have to be 
supplemented by internal rotation of qubits) 
the modular exponentiation part of Shor's algorithm \cite{Shor94} can be built almost 
exclusively from them.
Indeed
in detailed versions of this algorithm \cite{miquel96,knill96}
the most computer intensive part is the modular exponentiation.
This part can be built with essentially only {\tt T}-gates.
It goes without saying that the above 3-bit {\tt T}-gate can be turned into
a 7-bit one without much trouble combining the ideas of Figs 1-3.


Quantum logical gates presented here have the advantage that, 
at least ostensibly, they appear to use
the state of the control qubytes and act on the target qubytes as a whole.
One would expect this to improve the performance by allowing  one 
to correct for the errors which occur during the operation after 
it is already completed.
One way of verifying that this is indeed the case is to assume that errors existed in 
the individual qubites before the operation was carried out and to check if they can be 
still corrected for after the gate.  Below we give an example of how it works.

For the simplest case we look at the gate of Fig.1a.  
This gate works not only  for the state of 
Eq.(1) but also for the $|\pm_L\rangle$ discussed previously
(we are assuming now that if the control qubite is a + it flips the target bit).
For this case, it is rather transparent that the gate is indeed
a {\tt CNOT} on the qubytes.  The effect of decoherence on these states
is to flip the bites $|+\rangle$ to $|-\rangle$ and vice versa\cite{manlaf}.
If we assume that only 1 of the 6 qubits present in the {\tt CNOT} is affected
by decoherence it is possible to correct the final state even if the initial 
state was erroneous.  The first possibility is that initially the target qubyte
is incorrect. Fortunately, even if the qubyte is flipped 
the final syndrome is the same as the initial one.
Therefore for this case the state can be corrected as well before as after the gate.  If it is the control byte which has an erroneous bit, the error
propagates to the target bits. This error must then be corrected
before the next logical operation with these same qubytes as there are now
2 incorrect qubits. If the control
byte has a non-trivial syndrome, this will imply that the target byte
should also have the the same syndrome (if this is not true, there was
more than 1 error in the initial 6 bits).  

When the {\tt CNOT} is in the ($+,-$) basis, it is not possible to
do error correction between the individual  {\tt cnot}'s of pair of qubits
as the state is not in the code anymore.  If we perform a {\tt CNOT} in
the (0,1) basis as in Fig.1a-c, it is however possible to perform
intermediate error correction as the state remains in the code.
However in this case, the errors propagate differently.
For the case of Fig.1a when
the error is in the target bit we get an overall conditional sign flip
if the control qubyte is $|1_L\rangle$. This can be corrected by first checking
that the control byte is intact, then finding the syndrome of the target.
In addition to the appropriate 1-qubyte unitary transform, an overall sign flip
must be  performed conditionally to the control byte.
If the control byte has an error, we get a simpler answer.  In this case the 
error does not propagate to the target bit and can be corrected as if it would have been a  memory byte.

The case of the {\tt T}-gate is slighlty more involved.  In the (0,1) basis, the correction
of the error on the target byte will be conditional on the state of the 
control bytes.  If the error is in the control byte they
propagate through the logical operation without affecting the target byte.
These errors can be corrected after the operation or even before the next {\tt T}-gate.  The detailed behavior of the other erros will be discussed
elsewhere\cite{zulaf,pacs96}.

Obviously, this analysis can be generalised to the 5 and 7-bit codes
with the added complication of taking care of more bits and more types 
of errors\cite{zulaf,pacs96}.
But the main point is that it is possible to recover from errors even when 
logical operations are performed on erroneous qubytes, or if individual 
operations themselves
contaminate qubytes. We already know that different qubit-level designs of 
these and other qubyte
logical gates will have quite different error propagation properties. 
We expect that the choice of a particular design will depend on the specific 
physical implementation, and can be adjusted to minimize the effect of the most 
likely hardware problems. 

We would like to thank E.  Knill and B. Schumacher for many useful comments 
concerning classical and quantum error correction codes, and Rolf Landauer
for persistently asking the right question.

%

{\ }


{\ }
\vskip 0.5truein

Figure 1a. {\tt CNOT} (controlled-not) operation on encoded qubits.
Works in the (0,1) and ($+,-$) basis (see Eqs. (1) \& (3)).

\medskip

Figure 1b. Different implementation of {\tt CNOT}. Works only in the (0,1)
basis of Eq. (1). Corrections can be carried out inbetween individual 
{\tt cnot}'s, as the target qubyte (and, obviously, the control qubyte) 
are ``in the code'' after each {\tt cnot}. (This is also true for the 
(0,1) version of Fig. 1a.)

\medskip

Figure 1c. One more alternative of {\tt CNOT} in the (0,1) basis. 
The sequence of open dots connecting individual qubits of the control 
qubyte performs an {\tt XOR} (addition modulo 2). This version sums up 
the essence of the three bit {\tt CNOT}: It is enough to flip one of 
the target qubits depending on the parity of the three control qubits.
It may be also easy to implement in linear trap quantum computer\cite{Cirac95}.

\bigskip

Figure 2. {\tt CNOT} for a 7 bit code. This is only one of the several possible 
versions, in direct correspondence to Fig. 1a.

\bigskip

Figure 3a. Toffoli gate on encoded qubits (code of Eq. (1)). The operation 
$V$ is the ``square root'' of the {\tt cnot} (see text). This design which
protects only against decoherence can be obviously generalised to the 7-bit
code of Eq. (2) to protect against general 1-bit errors.

\medskip

Figure 3b. {\tt T}-gate for the ($+,-$) basis, Eq. (3). As before, 
other versions exist. 

\newpage
 
\input epsf.tex

\epsfxsize 2.8truein\epsfbox{ecnot3}

Figure 1a.
\vskip 0.5truein 

\epsfxsize 2.8truein\epsfbox{ecnot3c}

Figure 1b.
\vskip 0.5truein 

\epsfxsize 2.8truein\epsfbox{ecnot3d}

Figure 1c.
 
\vfill\eject

\epsfxsize 2.8truein\epsfbox{ecnot7}

Figure 2.
\vskip 0.2 truein

\epsfxsize 2.8truein\epsfbox{tof3b}

\vskip 0.6 truein Figure 3a.
\vskip 0.4 truein

\epsfxsize 2.8truein\epsfbox{tof3}

\vskip 0.8 truein  Figure 3b.
\vskip 0.4 truein

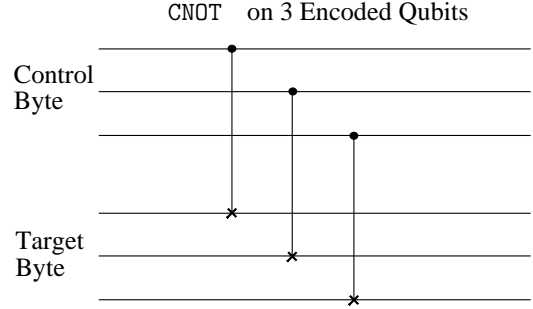
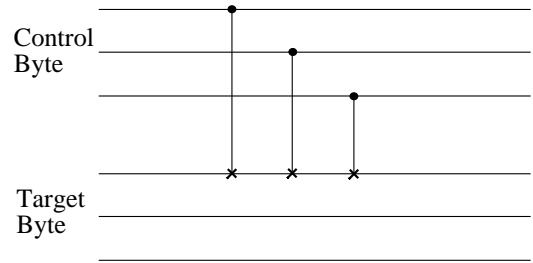
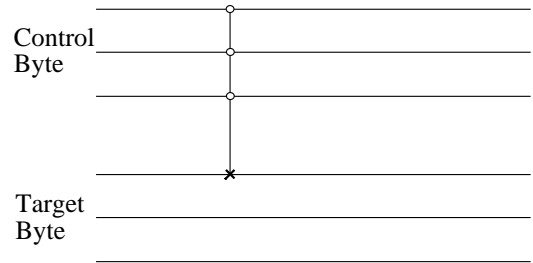
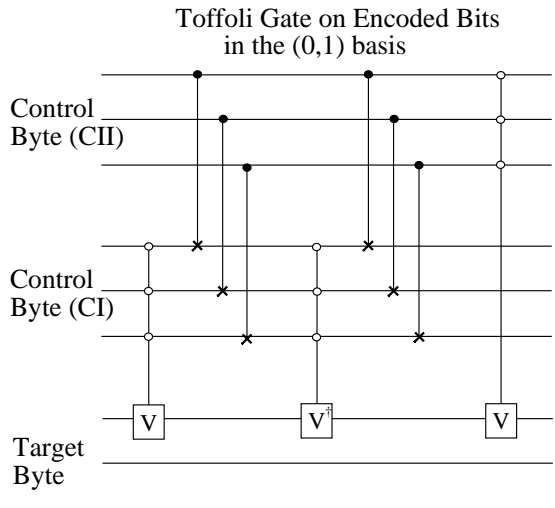
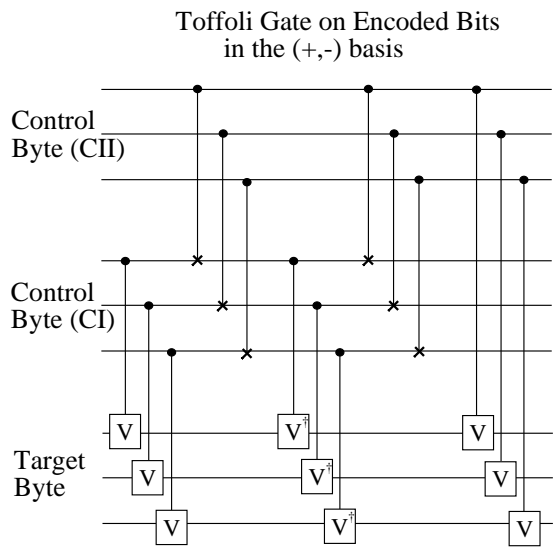
\end{document}